# Theoretical Limits of Joint Detection and Estimation for Radar Target

Nan Wang and Dazhuan Xu

College of Electronic and Information Engineering,

Nanjing University of Aeronautics and Astronautics, Nanjing, China

**Abstract:** This paper proposes a joint detection and estimation (JDE) scheme based on mutual information for the radar work, whose goal is to choose the true one between target existent and target absence, and to estimate the unknown distance parameter when the target is existent. Inspired by the thoughts of Shannon information theory, the JDE system model is established in the presence of complex white Gaussian noise. We make several main contributions: (1) the equivalent JDE channel and the posterior probability density function are derived based on the priori statistical characteristic of the noise, target scattering and joint target parameter; (2) the performance of the JDE system is measured by the joint entropy deviation and the joint information that is defined as the mutual information between received signal and the joint target parameter; (3) the sampling a posterior probability and cascaded JDEers[①] are proposed, and their performance is measured by the empirical joint entropy deviation the empirical joint information; (4) the joint theorem is proved that the joint information is the available limit of the overall performance, that is, the joint information is available, and the empirical joint information of any JDEer is no greater than the joint information; (5) the cascaded theorem is proved that the sum of empirical detection information and empirical estimation information can approximate the joint information, i.e., the performance limit of cascaded JDEer is available. Simulation results verify the correctness of the joint and the cascaded theorems, and show that the performance of the sampling a posterior probability JDEer is asymptotically optimal. Moreover, the performance of cascaded JDEer can approximate the system performance of JDE system.

## 1. Introduction

The JDE problems appear frequently in a diverse practical requirement, such as wireless communication, power systems, image processing and radar systems, and therefore have attracted much attention in recent years. Specifically, it is essential to first detect whether the object surface has a defect and estimate the shape in defect detection for image processing [1]. It is important to detect the correct topological model and estimate the system state in power grid monitoring[2]. In radar work [3], the existence or absence of the target is forced to determine and an estimate of the target position and associated parameters is given.

The most straightforward treatment is to handle the two subproblems separately using the corresponding

---

[①] JDEer is similar to the detector or estimator, but JDEer is used to deal with the JDE problems.

optimal methods for detection and estimation. For example, NP detection methods and Bayesian methods are used for detection and parameter estimation, respectively. However, this operation may not lead to an overall optimal performance, since detection and estimation are a symbiotic relationship in practical applications such as the above that can be met only by maximizing the overall optimal performance, and the detection results largely affect the estimated performance. Another equally straightforward approach is to deal with the joint problem with generalized likelihood ratio test [4, 5], which turns the complex likelihood ratio test into a simple hypothesis testing problem by obtaining the maximum likelihood estimate of the unknown parameter in each hypothesis. However, this method is optimal only when the number of samples is infinite, thus limiting the performance of the system in many cases. In essence, GLRT detects and then estimates, so the method achieves the corresponding estimated performance with optimal detection performance, rather than optimal at the same time.

The first method that treats the JDE problem in establishing a connection between the both can be traced back to 1968, that is, a linear combination of the detector and estimator in a Bayesian context investigated by Middleton and Esposito[6]. Later, [7] extends this work to the multiple-hypothesis testing case. The generalization to Bayesian risk can be handled by an integrated approach that relies on iterative algorithms to obtain the optimal solution[8]. The performance of the system in this framework is achieved by minimizing the overall Bayesian risk consisting of the probability of detection error and the estimated cost. However, it may not make practical physical sense to unify detection and estimation in a framework by simple linear combination.

G. V. Moustakides [9, 10] combined the NP like method with Bayesian estimation to deal with the joint problem and proposed one-step detection and two-step detection schemes and evaluated the joint performance using the average cost function that relies on both detection and estimation. In addition to the constraint on the false alarm probability in the NP method, the one-step detection scheme is added with a constraint on the probability of missing detection, and the two-step detection scheme is further strengthened with the constraint of reliable estimation. Based on it, a more general formula of JDE is presented in literature [11], and a decision dependent estimation cost is proposed to evaluate the overall performance. However, these methods optimizing the estimation performance subject to constraints on the detection performance. [12,13], in the sense of minimax, extends the analysis to composite hypothesis testing in [9,10,11]. However, it means that there are no unknown parameters to be estimated when the target does not exist for the radar problem. A worst-case cost function is proposed and the lower bound on the joint performance is given [14], but the unknown parameters are non-random and finite discrete sets.

In the field of radar, the application of Shannon's information theory to the radar problem can be traced back to the 1950s, when Woodward and Davies used the inverse probability principle to study the distance information problem [15]. In 1988, Bell used mutual information measure for waveform design [16] to study scattered information in spatial information. Later, the mutual information approach to solve radar problems has been widely used. In [17], the detection information that is defined as the mutual information between received signal and existent state, sampling a posteriori probability detector evaluated by empirical detection information is proposed. In addition, the radar target detection theorem is proved that the detection information is achievable, i.e., detection information is the maximum of empirical detection information. In [18], the joint distance-scattering information and entropy error are defined, and sampling a posterior probability estimator evaluated by empirical distance information is proposed. Moreover, the radar parameter estimation theorem is proved that the estimation information is achievable, i.e., the maximum of the empirical estimation information is the estimation information.

In this work, the problem of radar JDE is studied in the presence of complex white Gaussian noise from the perspective of mutual information for the radar application scenarios. Firstly, the radar JDE system model is established and whose performance is measured by defining joint information and joint entropy deviation. The sampling a posterior probability(SAP) JDEer and the cascaded JDEer are proposed, and their performance is measured by the empirical joint information and empirical joint entropy deviation. The joint theorem is proved that the joint information and joint entropy deviation are the available performance limit of the SAP JDEer. In addition, the cascaded theorem is proved that the sum of empirical detection information and empirical estimation information can approximate the joint information.

The remainder of the paper is organized as follows. Section 2 introduces the radar system model of joint target detection and parameter estimation. We define the radar information, entropy error and entropy deviation in Section 3 and define the sampling a posterior probability radar detector in Section 4. Sections 5 and Sections 6 prove the target and separation theorems, respectively. Sections 7 and provides the concluding remarks.

*Notations*: We use lower case letters to signify variable, upper case letters to denote random variable, bold italic lower case letters to signify column vectors and bold italic upper case letters to denote random column vectors. $E(\bullet)$ denotes the expectation operator. The operators $(\bullet)^T$ and $(\bullet)^H$ denote transpose and conjugate transpose of a vector respectively. A gaussian variable with expectation a and variance $\sigma^2$ is denoted by $N(a,\sigma^2)$. $I_0[\bullet]$ stands for the zero-order modified Bessel function of the first kind. $\pi(\bullet)$ denotes the priori probability. $d(\bullet)$ denotes the detection function. $diag(\bullet)$ denotes the diagonal matrix.

## 2. System Model

Consider a radar system with the transmitted baseband signals $\psi(t)$ is limited bandwidth $B/2$. Suppose that possible $K$ targets are completely contained in the observation interval, the observed signal is given by

$$y(t) = \sum_{k=1}^{K} v_k s_k \psi(t - \tau_k) + w(t). \tag{1}$$

Where $v_k \in \{0,1\}$ is the discrete variable of existence state, $v_k = 0$ and $v_k = 1$ corresponding to null-hypothesis $H_0$ and alternative hypothesis $H_1$ in a binary hypothesis testing, respectively. $s_k$ is the complex reflection coefficient, $\tau_k$ is the propagation delay. $w(t)$ is the complex additive white Gaussian noise process of limited bandwidth $B/2$ with zero mean and $N_0/2$ variance in its real and imaginary parts, respectively.

Assume that sample the received signal at the rate of $B$ samples, and correspondingly, $T$ is the sampling time. Define the observed sample number $N$ as the time-bandwidth product $N = TB$. The discrete-time equivalent of the received signal is given by

$$y(n) = \sum_{k=1}^{K} v_k s_k \psi(n - x_k) + w(n), n = -\frac{N}{2}, \cdots, \frac{N}{2} - 1. \tag{2}$$

Where $x_k = B\tau_k$ is the regularization delay. Due to the sample values are independent complex gaussian random variables, so are the real part and the imaginary part each other.

For briefness, (2) can be expressed as

$$\mathbf{y} = \mathbf{U}(x) diag(\mathbf{v})\mathbf{s} + \mathbf{w}. \tag{3}$$

Where $\mathbf{v} = (v_1, v_2, \cdots, v_K)$ is the existence state vector, $\mathbf{U}(x) = [\mathbf{u}(x_1), \mathbf{u}(x_2), \cdots \mathbf{u}(x_K)]$ is the position matrix of $K$ targets, $\mathbf{u}^T(x_k) = (\cdots, \psi(n - x_k), \cdots)$ is the sampling waveform of the $k$-th target. Equation (4) is a system model containing the existence state variable and target position parameter.

The probability density function (PDF) of the $N$-dimensional complex Gaussian noise vector $\mathbf{w}$ is given as

$$p(\mathbf{w}) = \left(\frac{1}{\pi N_0}\right)^N \exp\left(-\frac{1}{N_0}\|\mathbf{w}\|^2\right). \tag{5}$$

Given the existence state $\mathbf{v}$, the regularization delay vector $\mathbf{x}$ and the scattering vector $\mathbf{s}$, the multidimensional conditional PDF of $\mathbf{y}$ is obtained as

$$p(\mathbf{y}|\mathbf{v}, \mathbf{x}, \mathbf{s}) = \left(\frac{1}{\pi N_0}\right)^N \exp\left(-\frac{1}{N_0}\|\mathbf{y} - \mathbf{U}(x) diag(\mathbf{v})\mathbf{s}\|^2\right). \tag{6}$$

The constant module is a typical non-fluctuation, also known as the Swerling 0, in which the complex scattering coefficient is constant and the phases are distributed uniformly in $[0, 2\pi]$. Since the performance measure of multi-target for the constant module is very complicated, we only consider the case of a single target in the observed interval. The complex scattering coefficient is expressed as $s = \alpha e^{j\varphi}$, where $\alpha$ denotes the constant modulus and $\varphi$ denotes the phase. Suppose that the target position is also distributed uniformly within the observed interval $[-N/2, N/2]$ since a receiver has no the priori information about a target. Averaging over the random phase, the conditional PDF is given by

$$p(\mathbf{y}|v,x) = \left(\frac{1}{\pi N_0}\right)^N \exp\left[-\frac{1}{N_0}\left(\mathbf{y}^H\mathbf{y} + v\alpha^2\right)\right] I_0\left(\frac{2v\alpha}{N_0}\left|\mathbf{u}^H(x)\mathbf{y}\right|\right). \tag{6}$$

Where $\mathbf{u}^H(x)\mathbf{y}$ denotes the output of the matched filter. Equation (6) depicts the joint channel of target detection and parameter estimation. Since the random phase are averaged, the parameter estimate is specific to the target position in this paper.

## 3. Performance Measure

In this section, the posterior PDF of the joint target parameter is derived firstly. Then the joint information, the joint entropy error and the joint entropy deviation are defined based on the posterior PDF. In addition, the entropy number is defined for discrete source. Where the joint entropy error, the joint entropy deviation and the entropy number are negative indicators and the joint information is a forward indicator. The definition of performance indicators can unify the detector and estimator in a framework.

## 3.1 Posteriori Probability Density Function

Based on the Bayes formula and (6), the posteriori PDF of the joint target parameter $(v, x)$ after the received signal is derived as

$$p(v,x|\mathbf{y}) = \frac{1}{\upsilon}\pi(v,x)\exp(-v\rho^2) I_0\left(\frac{2v\alpha}{N_0}\left|\mathbf{u}^H(x)\mathbf{y}\right|\right). \tag{7}$$

Where

$$\upsilon = \sum_v \exp(-v\rho^2)\int_{-N/2}^{N/2} \pi(v,x) I_0\left(\frac{2v\alpha}{N_0}\left|\mathbf{u}^H(x)\mathbf{y}\right|\right) dx \tag{8}$$

denotes the regularization constant and $\rho^2 = \alpha^2/N_0$ denotes the signal to noise ratio(SNR). Generally, the prior distributions of existence state and position parameter of the target are independent of each other, i.e., $\pi(v,x) = \pi(v)\pi(x)$. The posteriori PDF $p(v,x|\mathbf{y})$ is also obtained by $p(x|v,\mathbf{y})$ and $\pi(v)$, i.e., $p(v,x|\mathbf{y}) = \pi(v)p(x|v,\mathbf{y})$. $p(x|v,\mathbf{y})$ denotes the posteriori PDF of position parameter given the existence state and received signal.

For a specific snapshot, assuming that $(v_0, x_0)$ is an actual joint target parameter, the received signal is given by

$$\mathbf{y}(n) = v_0 \alpha e^{j\varphi_0} \psi(n - x_0) + \mathbf{w}_0(n). \tag{9}$$

Substituting (9) into (7), one has

$$p(v,x|y) = \frac{1}{\upsilon}\pi(v,x)\exp(-v\rho^2)I_0\left(2v\rho^2\left|\sum_{n=-N/2}^{N/2}\left(v_0\psi(n-x)\psi(n-x_0)+\frac{1}{\alpha}e^{-j\varphi_0}\psi(n-x)w_0(n)\right)\right|\right). \quad (10)$$

Where the autocorrelation function $\sum_{n=-N/2}^{N/2}\psi(n-x)\psi(n-x_0)$ of the baseband signals is $\text{sinc}(x-x_0)$. For briefness, let the transmitted signal is $\psi(x)=\text{sinc}(x)$, the posterior PDF is given by

$$p(v,x|y) = \frac{1}{\upsilon}\pi(v,x)\exp(-v\rho^2)I_0\left(2v\rho^2\left|v_0\text{sinc}(x-x_0)+\frac{1}{\alpha}e^{-j\varphi_0}w_0(x)\right|\right). \quad (11)$$

Where $w_0(x) = \sum_{n=-N/2}^{N/2}\text{sinc}(n-x)w_0(n)$ is the complex additive white Gaussian noise process with zero mean and $N_0/2$ variance in its real and imaginary parts, respectively. $\frac{1}{\alpha}e^{-j\varphi_0}w_0(x)$ is also complex Gaussian noise process with zero mean and $1/\rho^2$ variance.

Equation (11) shows that the posterior PDF $p(v,x|y)$ of the joint target parameter given the received signal is stochastic since the $w_0(x)$ is different for each snapshot. However, the shape of posterior PDF is constant and is a symmetric distribution centered on the actual position $x_0$. Therefore, the posterior PDF curve is shifted along the target position changes.

## 3.2 Joint Information

The performance of the JDE system is measured by joint information that is short for the radar joint target detection and position estimation information. The information amount about the target existent state and position parameter from the received signal is contained in the joint information.

**Definition 1.** *Suppose $V$ is the existent state vector and $X$ is the position vector, $(V,X)$ is called the joint target parameter vector and $Y$ is the received signal vector. The mutual information $I(Y;VX)$ between the target state and the received signal is defined as the joint information, i.e.,*

$$I(Y;VX) = h(VX) - h(VX|Y). \quad (12)$$

*Where*

$$h(VX) = -\sum_v\sum_x \pi(vx)\log\pi(vx) \quad (13)$$

*is the priori entropy of the joint target parameter, and*

$$h(VX|Y) = E_y\left[-\sum_v\sum_x p(vx|y)\log p(xv|y)\right] \quad (14)$$

*is the posteriori entropy of the joint target parameter.*

According to the chain rule for information, Equation (12) can be rewritten as

$$I(Y;VX) = I(Y;V) + I(Y;X/V). \tag{15}$$

Where $I(Y;V)$ is the detection information[20], and $I(Y;X/V)$ is the estimation information[21] with known existence state. It shows that the joint information is the sum of the detection information and the estimation information, so that the performance of the JDE system can be evaluated uniformly by the amount of information for the both target detection and parameter estimation. Equation (15) also reveals the method of radar JDE, i.e., the target detection is first performed to obtain the detection information $I(Y;V)$, and then the parameter estimation is performed after the existence state is known to obtain the estimated information $I(Y;X/V)$.

## 3.3 Negative Indicators

Joint information is a forward measure indicator, and the overall performance gets progressively better as the joint information increases. In order to more comprehensively evaluate the performance of the JDE system, negative evaluation indicators are essential.

**Definition 2.** *Suppose $V$ is the existent state vector and $\mathbf{X}$ is the position vector, $Y$ is the received signal vector, $(\mathbf{V}, \mathbf{X})$ is called the joint target parameter vector with posterior entropy $h(\mathbf{VX}|\mathbf{Y})$. The joint entropy error of $(\mathbf{V}, \mathbf{X})$ is defined as*

$$\sigma_{EE}^2(VX|Y) = \frac{1}{2\pi e} 2^{2h(VX|Y)}. \tag{16}$$

From the posterior entropy formula, one has

$$\begin{aligned}\sigma_{EE}^2(VX|Y) &= 2^{2H(V|Y)} \cdot \frac{1}{2\pi e} 2^{2h(X/YV)} \\ &= \sigma_{EE}^2(V|Y) \cdot \sigma_{EE}^2(X/YV),\end{aligned} \tag{17}$$

where $\sigma_{EE}^2(V/Y) = 2^{2H(V/Y)}$ and $\sigma_{EE}^2(X/YV) = \frac{1}{2\pi e} 2^{2h(X/YV)}$ are the entropy error of the target detection and the parameter estimation systems, respectively. It can be seen that the joint entropy error of the JDE system is the product of the both. Similarly, the joint entropy deviation is given by

$$\sigma_{EE}(VX/Y) = \sigma_{EE}(V/Y) \cdot \sigma_{EE}(X/YV). \tag{18}$$

where $\sigma_{EE}(V/Y) = 2^{H(V/Y)}$ and $\sigma_{EE}(X/YV) = \frac{1}{\sqrt{2\pi e}} 2^{h(X/YV)}$ are the entropy deviation of the target detection and the parameter estimation systems, respectively. It can be seen that the joint entropy deviation of the JDE system is also the product of the both.

The system performance of the continuous distribution is measured by the entropy error and the entropy deviation, while the existence state $v$ is a discrete variable. Therefore, a new performance indicator is introduced in this paper for discrete random variables in order to analyze the physical significance of its entropy deviation.

**Definition 3.** *Suppose V is an element in a discrete set with the entropy* $H(V)$, *whose entropy number is defined as*

$$n_{EN} = 2^{H(V)}. \tag{19}$$

The entropy number indicates the effective number of elements in a discrete random variable, the entropy number is maximum when the variables are equal probability. For example, the entropy of n-ary equal probability source is $\log_2 n$, and whose entropy number is $n_{EN} = n$; the entropy of n-ary deterministic source is 0, and whose entropy number is $n_{EN} = 1$; the source $V \in \{1,2,3,\cdots\}$ with the probability distribution $P(V=i) = \frac{1}{2^i}$, whose entropy is $H(V) = 2$ and entropy number is $n_{EN} = 1$. Although there are an infinite number of elements in the set $V \in \{1,2,3,\cdots\}$, the entropy number is a finite value due to the elements are not equal probability in a set. In general, $1 \leq n_{EN} \leq Card$, where *Card* is the cardinal of a set, i.e., the number of elements in a set.

In the target detection, the entropy deviation of the target detector is defined as the entropy number of its posterior entropy. Detection performance increases with decreasing entropy number. If the entropy deviation is 1, it indicates that there is no uncertainty. The definitions of radar joint entropy error and radar joint entropy deviation of the parameter estimation system are consistent with the literature [21], where the former is a generalization of the mean square error and degenerates to the mean square error in the high SNR region.

## 4. Joint Detection and Estimation Schemes

We denote a target parameter as $\boldsymbol{\Gamma} = (V, X) \in \mathbb{N} \times \mathbb{R}$, and $\gamma = (v, x)$ is a sample(实现) in $\boldsymbol{\Gamma}$. A radar JDE system consists of two parts, the JDE channel and the JDEer, as shown in Fig. 1. The JDE channel is a conditional PDF $p(\mathbf{y}|v,x)$ from the joint target parameter space $\mathbb{N} \times \mathbb{R}$ to the $N$ dimensional complex space $\mathbb{C}^N$, whose input is the joint target parameter $\gamma = (v, x)$ and output is the received signal vector $\mathbf{y}$. The JDE channel is depicted by the conditional PDF of (6) and depends on the statistical properties of the radar systematic parameters and noise. The JDEer makes an estimation of the existence state and position parameter, which usually requires the use of the a priori statistical properties of the target.

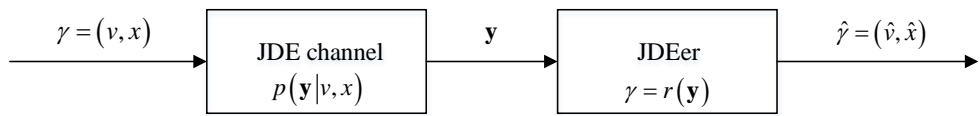

Fig.1 The JDE system

(1) Maximum A Posterior Probability JDEer

The maximum a posterior probability(MAP) JDEer obtains the decision that maximizes the posterior probability distribution function $p(v,x|\mathbf{y})$, i.e.

$$(\hat{v},\hat{x})_{\text{MAP}} = \arg\max_{v,x} p(v,x|\mathbf{y}), \tag{20}$$

where $p(v,x|\mathbf{y}) = p(\mathbf{y}|v,x)\pi(x)\pi(v)/p(\mathbf{y})$. The maximum likelihood JDEer can be used if the priori probabilities of the joint target parameters are equal. The maximum likelihood radar JDEer obtains the decision that maximizes the likelihood function $p(\mathbf{y}|v,x)$, i.e., $(\hat{v},\hat{x})_{\text{ML}} = \arg\max_{v,x} p(\mathbf{y}|v,x)$. But, in general, the probability of target existence or absence and the probability that the target is located at each point in the observation interval are likely to be unequal. Hence, the maximum likelihood method is not applicable to the design of actual radar JDEer.

(2) Sampling A Posterior Probability JDEer

The decision of SAP JDEer is corresponding to the "lots" drawn from a set consist of the posterior probability of existence states. Therefore, SAP JDEer is stochastic. Conversely, MAP JDEer is deterministic since they judge the target must be existent or not. In other words, for the joint target parameter, the decision of SAP JDEer may be 0.2 or 0.7, while the that of MAP JDEer can only be 0 or 1. Thus, the deterministic method is relatively crude since the target detection is essentially a probabilistic event. SAP radar JDEer is defined as following.

**Definition 4**. *Suppose a JDEer with the transmitted joint target parameter $(v,x)$, if its outputted is*

$$(\hat{v},\hat{x})_{\text{SAP}} = \arg\operatorname{smp}_{v,x} p(v,x|\mathbf{y}), \tag{21}$$

*this JDEer is called SAP JDEer. where $p(v,x|\mathbf{y})$ is the posteriori PDF of joint target parameter conditioned on the received signal, and $\arg\operatorname{smp}_{v,x} p(v,x|\mathbf{y})$ denotes the sampling operation.*

The decision results are uncertain given received signal since the stochastic feature of SAP JDEer and whose average performance depends on the posteriori PDF $p(v,x|\mathbf{y})$ of SAP JDEer, $p(v,x|\mathbf{y})$ is obtained by

$$p(\hat{v},\hat{x}|\mathbf{y}) = p(v,x|\mathbf{y}). \tag{22}$$

Therefore, the joint information of SAP JDEer is same as the corresponding theoretical value.

(3) Cascaded JDEer

The JDE process is divided into two stages, target detection and parameter estimation. In the first stage, the information about the existence state is obtained by the target detector from the received signal. In the second stage, an estimation of the target position is gained by the parameter estimator based on the detection result and the received signal. To obtain the optimal overall performance, the target detector and parameter estimator are cascaded to form the cascaded JDEer, as shown in fig. 2.

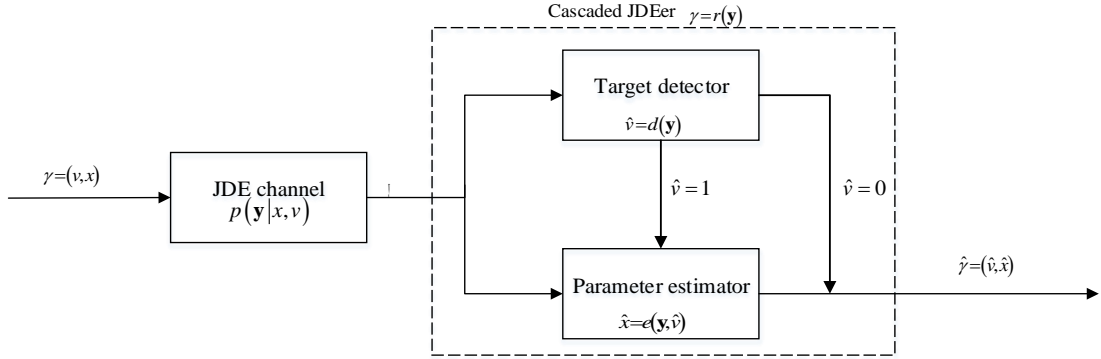

Fig.2 The cascaded JDE systems

One detection process of the cascaded system consists of the joint target parameter, JDE channel, target detector and parameter estimator. The estimator is directed by the result of the detection operation. Specifically, target detector and parameter estimator work simultaneously, and the estimator does not work unless the detector has decided that the signal is present. It is worth noting that the target detector outputs the existence state along with the probability of the corresponding state, which is equivalent to the soft judgment of the demodulator in the communication system.

**Target Detector**

According to the conditional probability distribution of detection channel (8), the posteriori probability distribution of the existent state conditioned on the received signal is given by[20]

$$P(v|\mathbf{y}) = \frac{1}{\upsilon}\pi(v)\exp(-v\rho^2)\frac{1}{N}\int_{-N/2}^{N/2} I_0\left(2v\rho^2\left|v\sin c(x-x_0)+\frac{1}{\alpha}w(x)\right|\right)dx, \qquad (23)$$

where

$$\upsilon = \sum_v \pi(v)\exp(-v\rho^2)\frac{1}{N}\int_{-N/2}^{N/2} I_0\left(2v\rho^2\left|v\sin c(x-x_0)+\frac{1}{\alpha}w(x)\right|\right)dx. \qquad (24)$$

The detection information of the target detector can be calculated based on the posterior probability distribution of the existent state.

**Parameter Estimator**

According to the conditional PDF of estimation channel (8), the posteriori PDF of the target position conditioned on the received signal and the existent state is given by

$$p(x|\mathbf{y},v) = \frac{1}{\upsilon}I_0\left(2v\rho^2\left|v_0\sin c(x-x_0)+\frac{1}{\alpha}w(x)\right|\right), \qquad (25)$$

where

$$\upsilon = \int_{-N/2}^{N/2} I_0\left(2v\rho^2\left|v_0\sin c(x-x_0)+\frac{1}{\alpha}w(x)\right|\right)dx. \qquad (26)$$

The estimation information and entropy error of the parameter estimator can be calculated based on (25).

Clearly, Equation (25) is consistent with the posteriori PDF in [21] when the target exists.

## 5. Radar Theorems

Two theorems are given in this section, namely the joint and the cascaded theorems that point out the performance limits of the SAP JDEer and cascaded JDEer.

## 5.1 Joint Theorem

The $m$ memoryless extended joint target parameters means that the extended joint target parameters $\pmb{\Gamma}^m$ are independent of each other, denoted as $\pmb{\Gamma}^m = (\pmb{V}^m, \pmb{X}^m) \in \mathbb{N} \times \mathbb{R}$. The JDEer is a function $\hat{\gamma} = r(\pmb{y})$ of the received signal, and $\hat{\gamma} = (\hat{v}, \hat{x})$ is an decision of the joint target parameter $\gamma = (v, x)$ made by the JDEer given the received sequence. One JDE process that consists of the extended joint target parameter, the extended JDE channel and the JDEer is denoted. The extended joint target parameter and the extended radar JDE channel are produced by multiple snapshots. The JDE process with $m$ snapshots is shown in fig. 3, and $(\gamma^m, \pmb{Y}^m, \hat{\gamma}^m)$ form a Markov chain obviously.

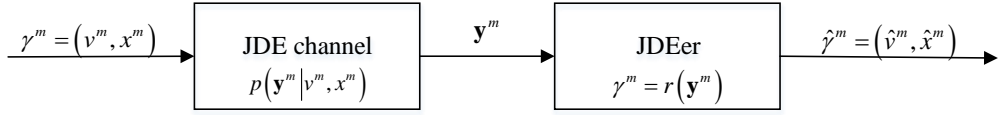

Fig.3 The JDE system with $m$ snapshots

Many basic definitions required for the subsequent development of the joint theorem are introduced. Given the joint target parameter and JDE channel, the empirical joint entropy and the empirical joint information of the SAP JDEer are defined below.

***Definition 4.*** *The empirical joint entropy, the empirical joint entropy error, the empirical joint entropy deviation and the empirical joint information obtained by the SAP JDEer from M snapshots are defined as*

$$h^{(M)}(\hat{\pmb{\Gamma}};Y) = -\frac{1}{M}\log p(\hat{\gamma}^M | \pmb{y}^M) \tag{27}$$

$$\sigma_{EE}^{2(M)}(\hat{\pmb{\Gamma}};Y) = \frac{1}{2\pi e} 2^{-\frac{2}{M}\log p(\hat{\gamma}^M | \pmb{y}^M)} \tag{28}$$

$$\sigma_{EE}^{(M)}(\hat{\pmb{\Gamma}};Y) = \frac{1}{\sqrt{2\pi e}} 2^{-\frac{1}{M}\log p(\hat{\gamma}^M | \pmb{y}^M)} \tag{29}$$

$$I^{(M)}(\hat{\pmb{\Gamma}};Y) = \frac{1}{M}\log \frac{p(\hat{\gamma}^M | \pmb{y}^M)}{p(\hat{\gamma}^M)} \tag{30}$$

***Definition 5.*** *The joint information $I(\pmb{\Gamma};Y)$ is said to be available if there exists a JDEer with empirical joint*

information $I^{(M)}(\hat{\boldsymbol{\varGamma}};Y)$ for $M$ snapshots satisfy $\lim_{M\to\infty} I^{(M)}(\hat{\boldsymbol{\varGamma}};Y) = I(\boldsymbol{\varGamma};Y)$. Similarly, the joint entropy error $\sigma_{EE}^2(\boldsymbol{\varGamma}|Y)$ is said to be available if there exists a JDEer with empirical joint entropy error $\sigma_{EE}^{2(M)}(\hat{\boldsymbol{\varGamma}}|Y)$ for $M$ snapshots satisfy $\lim_{M\to\infty}\sigma_{EE}^{2(M)}(\hat{\boldsymbol{\varGamma}}|Y) = \sigma_{EE}^2(\boldsymbol{\varGamma}|Y)$.

**Definition 6.** The set $A_\varepsilon^{(M)} = \{(\boldsymbol{\gamma}^M, \boldsymbol{y}^M) \in \Gamma^M \times \mathbb{R}^M\}$ of $\varepsilon$-jointly typical sequence $\{(\boldsymbol{\gamma}^M, \boldsymbol{y}^M)\}$ with respect to the distribution $p(\boldsymbol{\gamma}, \boldsymbol{y})$ is a subset of the set of M-sequences with empirical entropy $\varepsilon$-close to the true entropies, i.e.

$$\left|-\frac{1}{M}\log P(\boldsymbol{\gamma}^M) - h(\boldsymbol{\varGamma})\right| < \varepsilon \tag{31}$$

$$\left|-\frac{1}{M}\log p(\boldsymbol{y}^M) - h(Y)\right| < \varepsilon \tag{32}$$

$$\left|-\frac{1}{M}\log p(\boldsymbol{\gamma}^M, \boldsymbol{y}^M) - h(\boldsymbol{\varGamma},Y)\right| < \varepsilon\} \tag{33}$$

where

$$p(\boldsymbol{\gamma}^M, \boldsymbol{y}^M) = \prod_{m=1}^{M} p(\boldsymbol{\gamma}_m, \boldsymbol{y}_m) \tag{34}$$

**Lemma 1.** For a memoryless snapshot channel $p(\boldsymbol{y}^M|\boldsymbol{\gamma}^M)$, if $\hat{\boldsymbol{\gamma}}^M$ is a simple of independent $M$-samples of the $p(\boldsymbol{\gamma}|\boldsymbol{y})$, $(\hat{\boldsymbol{\gamma}}^M, \boldsymbol{y}^M)$ are jointly typical sequences with respect to the $p(\hat{\boldsymbol{\gamma}}^M, \boldsymbol{y}^M)$.

*Proof:* Since $\hat{\boldsymbol{\gamma}}^M$ is a simple of independent $M$-samples by sampling the posteriori PDF $p(\boldsymbol{\gamma}|\boldsymbol{y})$, the posteriori PDF of SAP radar JDEer is $p_{SAP}(\hat{\boldsymbol{\gamma}}^M|\boldsymbol{y}^M) = p(\hat{\boldsymbol{\gamma}}^M|\boldsymbol{y}^M)$, one has

$$\begin{aligned} p_{SAP}(\hat{\boldsymbol{\gamma}}^M, \boldsymbol{y}^M) &= p(\boldsymbol{y}^M)p_{SAP}(\hat{\boldsymbol{\gamma}}^M|\boldsymbol{y}^M) \\ &= p(\boldsymbol{y}^M)p(\hat{\boldsymbol{\gamma}}^M|\boldsymbol{y}^M) \\ &= p(\hat{\boldsymbol{\gamma}}^M, \boldsymbol{y}^M) \end{aligned} \tag{35}$$

End of proof.

From Lemma 1 and the definition of the jointly typical sequences, one has

$$\left|-\frac{1}{M}\log p(\hat{\boldsymbol{\gamma}}^M) - h(\boldsymbol{\varGamma})\right| < \varepsilon \tag{36}$$

**Theorem 1.** The joint information $I(\boldsymbol{\varGamma};Y)$ is available, specifically, if the statistical property of the joint target parameter and JDE channel are known by the JDEer, for every $\varepsilon > 0$, there exists the empirical joint information of the JDEer satisfies

$$I(Y;\boldsymbol{\varGamma}) - 3\varepsilon < I(Y^M;\boldsymbol{\varGamma}^M) < I(Y;\boldsymbol{\varGamma}) + 3\varepsilon, \tag{37}$$

$$\lim_{M\to\infty} I(Y^M;\boldsymbol{\varGamma}^M) = I(Y;\boldsymbol{\varGamma}). \tag{38}$$

*Conversely, the empirical joint information of any JDEer is no greater than the joint information.*

**Remark:** *As an achievable method of radar JDE, SAP JDEer, is asymptotically optimal. In other words, the maximum of empirical joint information obtained by SAP JDEer is joint information. The application of joint theorem is extensive and the proof is given only for a single target, but its framework of proof can be extended to multi-target, while it is not considered for the sake of brevity.*

## 5.2 Cascaded Theorem

The target detection theorem[20] and the parameter estimation theorem[21] are proved. When detection and estimation are considered as two independent processes, the corresponding SAP target detector and SAP parameter estimators can be used separately to achieve the limits of detection performance and estimation performance. The cascaded radar JDEer couples the detector to the estimator to obtain the overall optimal performance.

**Theorem 2**. *As the number of snapshots increases, the sum of the empirical detection information obtained by the SAP target detector and the empirical estimation information gained by the SAP estimator can approximate the limit of the joint information.*

The joint information is the performance limit of the cascaded JDEer that consist of SAP detector and SAP estimator, and detection information and estimation information can be obtained by increasing the number of snapshots, that is, the cascaded JDEer can obtain the performance limits of the JDEer system.

## 6. Numerical Results

In this section, numerical simulations are performed to evaluate the performances of proposed detectors, and to verify the radar and separation theorems. We consider the time bandwidth product $TB = 128$.

### 6.1. The Verification of Radar Theorem

The analytic and numerical curves of the radar information and the empirical radar information of the SAP radar detector for different snapshot numbers are compared in Fig. 4, in which, the empirical radar information gained by 10,000 snapshots, 1000 snapshots and 100 snapshots respectively. As seen, the empirical radar information gradually approaches the radar information as the number of snapshots increases, and it almost overlaps with the radar information for the 10,000 snapshots, which indicates that the radar information is achievable. In Fig.9, the empirical detection information, empirical estimation information and empirical radar information are approximate the corresponding theoretic limit, i.e., detection information, estimation information and radar information for the 10,000 snapshots. Similarly, the same is true for entropy deviation.

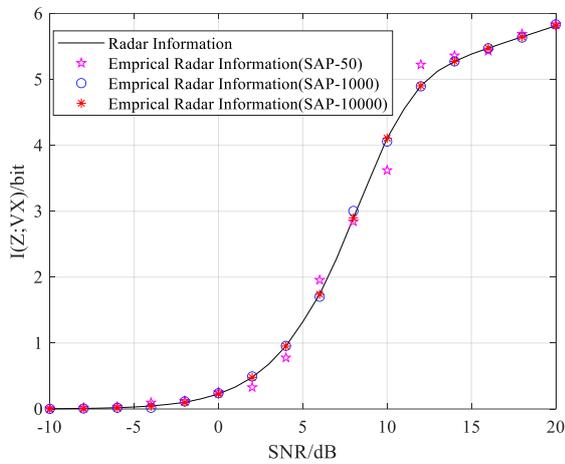 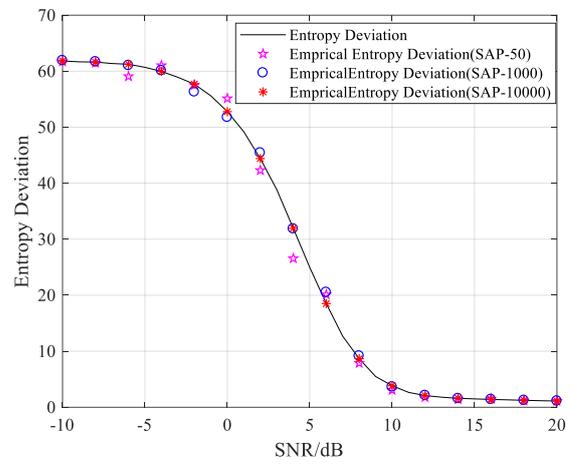

(a) Radar information and empirical radar information for various numbers

(b) Entropy deviation and empirical entropy deviation for various numbers

Fig.4 Radar information and entropy deviation of radar detector, and empirical radar information and empirical entropy deviation of coupling radar detector for different number.

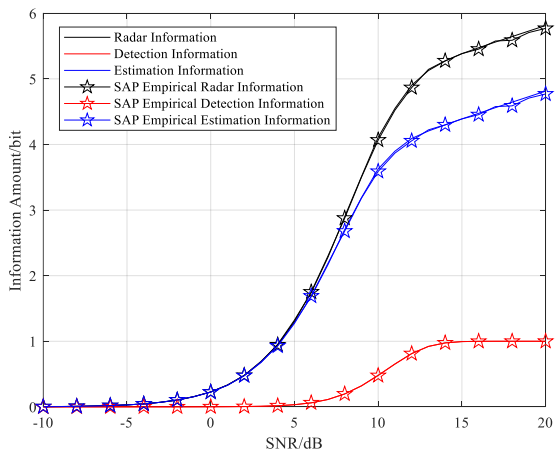 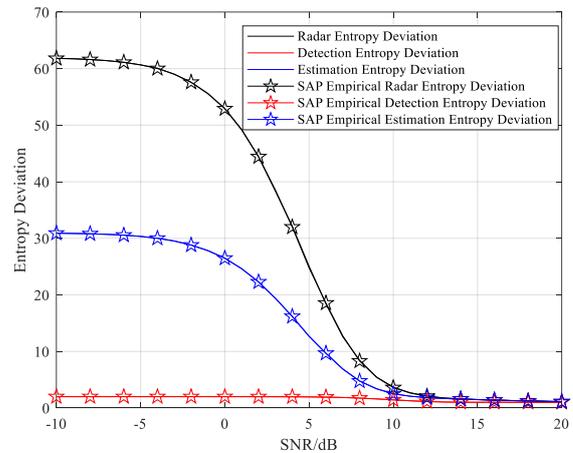

(a) Empirical information of SAP detector and information

(b) Empirical entropy deviation of SAP detector and entropy deviation

Fig.5 Information and entropy deviation of radar detector, and empirical information and empirical entropy deviation of coupling radar detector for different SNR(10,000 snapshots).

## 6.2 The Verification of Cascading Theorem

The radar information of the joint and the cascaded radar systems are shown in Fig. 6, which show that the radar information is the sum of the detection information and the estimation information.

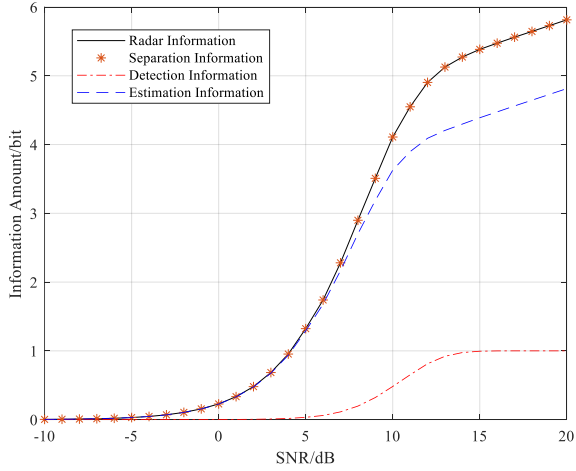 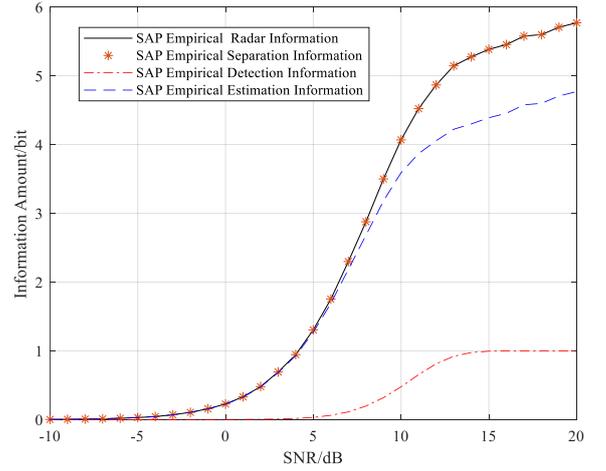

Fig. 6 The radar information and separation information of the coupling radar system, detection information of target detector and estimation information of parameter estimator.

Fig. 7 The empirical radar information and empirical separation information of the coupling radar system, empirical detection information of target detector and empirical estimation information of parameter estimator.

The empirical RI of SAP joint and concatenated target detector and parameter estimator is shown in Fig. 7. It can be seen that the RI obtained by both radar detectors is consistent, which verifies the correctness of the separation theorem of target detection and parameter estimation.

# 7. Conclusion and Further Work

We have formulated the RJDE measured by defined radar information and entropy deviation. The proposed SAP radar JDEer is measured by the empirical radar information and empirical entropy deviation. We prove the radar and cascading theorems that indicate the radar information is the achievable performance limit of the radar JDEer and the obtained from empirical detection information and empirical detection information by cascading radar JDEer.

# Appendix

*A. The proof of joint theorem*

*Proof:* The proof of theorem is constructive and divided into two parts, the positive to and the converse to theorem, and the former is proved first.

*Proof of the positive to theorem:*

The simple $\gamma^M$ of $M$ extended target sources are generated independently according to the priori distribution,

and the received sequence $y^M$ is generated based on the $\gamma^M$ and $M$ extended channel $p(y^M | \gamma^M)$, which satisfies

$$p(y^M | \gamma^M) = \prod_{m=1}^{M} p(y_m | \gamma_m). \tag{41}$$

Introducing SAP detector, let $\hat{\gamma}^M$ be the M-sampling estimation for memoryless snapshot channels $p(y^M | \gamma^M)$, then $(\hat{\gamma}^M, y^M)$ is a joint typical sequence about the probability distribution $p(\hat{\gamma}^M, y^M)$ according to Lemma 1.

Based on the definition of the joint typical sequence, for every $\varepsilon(\varepsilon > 0)$, as long as $M$ is large enough, then

$$\left| -\frac{1}{M} \log p(y^M) - h(Y) \right| < \varepsilon, \tag{42}$$

$$\left| -\frac{1}{M} \log p(\hat{\gamma}^M, y^M) - h(\Gamma, Y) \right| < \varepsilon, \tag{43}$$

Due to $p(\hat{\gamma}^M | y^M) = p(\hat{\gamma}^M, y^M) / p(y^M)$, then

$$\left| -\frac{1}{M} \log p(\hat{\gamma}^M | y^M) - h(\Gamma | Y) \right| < 2\varepsilon, \tag{44}$$

or

$$h(\Gamma|Y) - 2\varepsilon < -\frac{1}{M} \log p(\hat{\gamma}^M | y^M) < h(\Gamma|Y) + 2\varepsilon. \tag{45}$$

According to the definition of empirical radar information, one has

$$I(Y;\Gamma) - 3\varepsilon < I(Y^M; \Gamma^M) < I(Y;\Gamma) + 3\varepsilon \tag{46}$$

then according to Chebyshev law of large numbers, with the number of snapshots $M \to \infty$, one has

$$\lim_{M \to \infty} I(Y^M; \Gamma^M) = I(Y;\Gamma) \tag{47}$$

*Proof of the converse to theorem:*

Assume that $\hat{\gamma}^M = r(y^M)$ is arbitrary detector whose empirical radar information is denoted as $I(\gamma^M; \hat{\gamma}^M)$. $(\Gamma^M, Y^m, \hat{\Gamma}^M)$ forms a Markov chain from Fig.3, and according to the data processing theorem, one has

$$I(\hat{\Gamma}^M; \Gamma^M) \leq I(Y^M; \Gamma^M) = I(Y;\Gamma) \tag{50}$$

Then the empirical posterior entropy

$$h(\hat{\Gamma}^M; \Gamma^M) \geq h(Y;\Gamma) \tag{51}$$

Since the empirical entropy corresponds to the empirical entropy error, then the empirical entropy error of any radar detector is greater than the entropy error.

End of the proof. ∎

B. *The proof of cascaded theorem*

*Proof:* According to the additive property of the radar information, one has

$$I(Y;VX) = I(Y;V) + I(Y;X|V). \tag{52}$$

Based on the target detection theorem, the empirical detection information of the SAP target detector can approach infinitely the limit $I(Y;V)$. And

$$\begin{aligned} I(Y;X|V) &= \pi(0)I(Y;X|0) + \pi(1)I(Y;X|1) \\ &= \pi(1)I(Y;X|1) \end{aligned} \tag{53}$$

where $I(Y;X|1)$ is the estimation information with the target exists.

Substituting (53) into (52) yields

$$I(Y;VX) = I(Y;V) + \pi(1)I(Y;X|1). \tag{54}$$

According to the parameter estimation theorem, the empirical estimation information of the SAP estimator can approach infinitely the limit $I(Y;X|1)$. Therefore, the cascaded JDE system consists that the SAP target detector and the SAP estimator can approximate the limit of the joint information $I(Y;VX)$. ∎

End of Proof.